\documentclass[a4]{article}

\usepackage{graphicx}
\usepackage{amsmath}
\usepackage{amssymb}
\usepackage{enumerate}
\usepackage{xcolor}

\def\ben{\begin{enumerate}}
\def\een{\end{enumerate}}
\def\bit{\begin{itemize}}
\def\eit{\end{itemize}}
\def\beq{\begin{equation}}
\def\eeq{\end{equation}}
\def\bea{\begin{eqnarray}}
\def\eea{\end{eqnarray}}
\def\bq{\begin{quote}}
\def\eq{\end{quote}}
\def \lsim{\mathrel{\vcenter
     {\hbox{$<$}\nointerlineskip\hbox{$\sim$}}}}
\def \gsim{\mathrel{\vcenter
     {\hbox{$>$}\nointerlineskip\hbox{$\sim$}}}}
\def\gappeq{\mathrel{\rlap {\raise.5ex\hbox{$>$}}
{\lower.5ex\hbox{$\sim$}}}}
\def\lappeq{\mathrel{\rlap{\raise.5ex\hbox{$<$}}
{\lower.5ex\hbox{$\sim$}}}}

\def\Tmn{T^{\mu \nu}}

\def\a{\alpha}

\newcommand{\fPQ}{f_{\rm PQ}}
\newcommand{\mpl}{m_{\rm Pl}}
\newcommand{\LQCD}{\Lambda_{\rm QCD}}

\begin{document}

\renewcommand{\thefootnote}{\fnsymbol{footnote}}
\begin{center}
{\Large Rotating Drops of Axion  Dark Matter 
}

\vskip 25pt
{\bf   Sacha Davidson $^{1,}$\footnote{E-mail address:
s.davidson@ipnl.in2p3.fr} and Thomas Schwetz $^{2,}$}\footnote{E-mail address:
schwetz@kit.edu} 
 
\vskip 10pt  
$^1${\it 
Univ. Lyon, Universit\'e Lyon 1, CNRS/IN2P3, IPN-Lyon, F-69622, 
Villeurbanne, France 
}\\
$^2${\it Institut f\"ur Kernphysik, Karlsruher Institut f\"ur Technologie (KIT),
D-76021 Karlsruhe, Germany} \\
\vskip 20pt
{\bf Abstract}
\end{center}

\begin{quotation}
  {\noindent\small We consider how QCD axions produced by the
    misalignment mechanism could form galactic dark matter halos.  We
    recall that stationary, gravitationally stable axion field
    configurations have the size of an asteroid with masses of order
    $10^{-13} M_\odot$ (because gradient pressure is insufficient to
    support a larger object).  We call such field configurations
    ``drops''.  We explore whether rotating drops could be larger, and
    find that their mass could increase by a factor $\sim 10$.
    Remarkably this mass is comparable to the mass of miniclusters
    generated from misalignment axions in the scenario where the axion
    is born after inflation. We speculate that misalignment axions
    today are in the form of drops, contributing to dark matter like a
    distribution of asteroids (and not as a coherent oscillating
    background field). We consider some observational signatures of
    the drops, which seem consistent with a galactic halo made of
    axion dark matter.

\vskip 10pt
\noindent
}

\end{quotation}

\vskip 20pt

\renewcommand{\thefootnote}{\arabic{footnote}}
\setcounter{footnote}{0}

\section{Introduction}
\label{intro}

The QCD axion~\cite{rev,revggr} is a motivated, minimal and very
curious  dark matter candidate. It originally appeared~\cite{axion}
in Peccei and Quinn's  solution to the strong CP problem~\cite{PQ},
as the pseudo-goldstone boson of  a global, anomalous $U_{\rm PQ}(1)$ symmetry.
In  ``invisible'' axion models~\cite{russes,DFS} which
agree with observations, heavy new scalars and/or
fermions are introduced, and the  $U_{\rm PQ}(1)$  is
spontaneously broken at a high scale $\fPQ \sim 10^{9} \to 10^{11}$ GeV,
 so that the only new particle 
at accessible energies is the light, feebly-coupled axion. 
And despite that the axion and neutrino have comparable masses,
the axion is a cold dark matter candidate,
due to its non-thermal production in cosmology.

There are two production mechanisms for axion cold dark matter, in 
the case where the Peccei-Quinn phase transition occurs after inflation.
Both occur around the QCD phase transition, when the
axion mass turns on.
The ``misalignment mechanism'' \cite{DineFischler,PWW}, 
produces an  oscillating  classical axion field, 
and the decay of  the string network  produces a distribution
of  cold axion modes. The classical field, produced by the
misalignment mechanism, will be called ``the axion
field'' in the following. This may be what some authors
refer to as a Bose Einstein Condensate, however, the
literature is confusing because other authors discuss
whether the  misalignment axions could ``gravitationally
thermalise'',  as a prerequisite to forming a Bose
Einstein Condensate.

In this paper, we focus on the axion field, despite that
most axion dark matter may be produced by strings 
\cite{HKSS,KSS}. As pointed out by
Sikivie \cite{SY1},  dark matter composed of an axion field
is different from WIMPs. The $T^{ij}$ elements (pressures)
of the stress-energy tensor  are
different, and the axion field is single-valued,
whereas WIMPs are described by a phase space distribution. 
The difference is intuitive and clear during late-time structure
formation: particles (described by phase space)
fall into a gravitational well,  rise up the other
side, fall back, and so on. Indeed the whole population
of particles does this simultaneously, interacting only
via gravity in the usual CDM approximation. This
can be modelled via N-body simulations. The axion field,
being single-valued, is like a fluid, so its
velocity must remain single-valued as it falls in,
possibly leading to shocks and turbulence.

The original  aim of this project was to address the question ``How to make
the halo of Andromeda (or any other galaxy) with the QCD  axion field?''.
This is both a dynamical question, about the evolution
of axion dark matter from the QCD phase transition until
today, and  a ``stationary'' question, about
the axion field configuration in the galaxy over
the past several billion years. 
Sections \ref{secA} and \ref{sec:num} address the ``stationary''
part of the question.
We give approximate  solutions of the  equations
of motion of the axion and Newtonian gravity, 
and  call these   self-gravitating
solutions  ``drops''. Such solutions
have been studied for a wide range
of parameters,  in the literature
for  ``Bose stars''~\cite{bosestarrev,Jetzer:1991jr} and 
galactic halos~\cite{RDS,Chavanis}.
They are discussed very completely by Rindler-Daller
and Shapiro (RDS)~\cite{RDS} and Chavanis, Chavanis and Delfini~\cite{Chavanis},
where can be found extensive references. 
 As pointed out by Barranco and Bernal~\cite{BB}, and
noticed by several authors~\cite{RDS,Chavanis}, 
the mass and size of QCD axion  drops is $M\sim 10^{-13} M_\odot$
and $R \sim 100$~km, which is small compared to
Andromeda ($M\sim 10^{12} M_\odot$, $R \sim 100$ kpc)\footnote{Recall that
$M_\odot  = 1.1 \times 10^{57} ~{\rm GeV}~ = 2.0  \times 10^{30} ~{\rm kg}~  $,
kpc $= 3.086 \times 10^{21}$ cm.}.
Chavanis \cite{Chavanis} showed that the negative self-interaction of
QCD axions gives an upper bound of this order on
the size of a  non-rotating drop. We allow our
drops to rotate, and find by analytic estimates (section~\ref{secA})
as well as numerical solutions of the relevant equations (section~\ref{sec:num})
that rotating drops can be about one order of magnitude more massive.

Section \ref{secB} reviews the history of the Universe
in the scenario where the Peccei-Quinn phase transition occurs
after inflation. Hogan and Rees \cite{mini} noticed
that in this scenario,  there
are ${\cal O}(1)$ fluctuations in the 
energy density of the
axion field  on the scale of the  horizon at the QCD  
phase transition. 
The mass of
these fluctuations, referred to as ``miniclusters''
can be comparable or larger than   the axion drop mass. 
So  we use the virial theorem to hypothesize that,  
when  miniclusters  gravitationally
collapse, the larger ones  fragment into axion drops. 
(The collapse
of miniclusters should be addressed
numerically, which we hope to do shortly.)
If this is the case, then the
axion field dark matter in the Universe
today would be in the form of axion drops,
which would behave as WIMPs~\cite{BB}. In particuliar,
this implies that in the neighbourhood of
the sun,  any coherently
oscillating background  axion field is
small, and is not determined
by the dark matter density, with possibly severe
implications for axion dark matter searches \cite{ADMX}.

Section \ref{secC} reviews the constraints on dark matter in the form
of axion drops, which are within the mass range of ``macro dark
matter'', studied in \cite{macro}.  We summarize in
section~\ref{sum}. In the appendix we provide a derivation of the
non-relativistic field equations coupled to Newtonian gravity,
starting from the general relativistic action of a real scalar field.


\section{The axion ``drop'': a stable gravitationally bound configuration}
\label{secA}

The aim of this section is to identify 
stable  configurations of the QCD axion field,
in  the presence of self-interactions and Newtonian gravity.
We refer to these 
configurations as ``drops''\footnote{We
thank a seminar participant  at Zurich University for this name.}.
This question has been widely studied \cite{RDS,Chavanis,BB,Eby:2014fya,Eby:2015hsq}; 
our new contribution is to allow
the drops to rotate.  We
review  analytic estimates for the
mass and radius of the drops, 
as a function of the various microscopic
and/or external  parameters ($m, \fPQ, \mpl$,...),
which imply that 
the drops  ressemble asteroids.
The purpose of the analytic estimates
is to understand how the mass and
radius of the drops scale; 
 so the  estimates only need
to be of the right order of magnitude. 

The drops may rotate, but we
neglect time-variation of the
 radial density profile.
This means 
that our drop is not allowed a  ``breathing mode'',
which could be compatible with long-term
stability;  we suppose that this
would not significantly change the parameters
we are interested in.

The stress-energy tensor for
 the real QCD axion field $a$ is $  T^\mu_\nu= a^{;\mu} a_{;\nu} -
 \left[  \frac{1}{2} a^{;\a}
a_{;\a} - V(a)\right] \delta^\mu_\nu$,
where the potential  after the
QCD phase transition, is\footnote{We adopt in this paper the 
dilute-instanton approximation for the potential, which suffices for our purposes. { A recent discussion of some modifications can be found in \cite{Villadoro}.}}
\beq
V(a) \approx f^2_{\rm PQ}m^2 [1-\cos(a/\fPQ)] \simeq \frac{1}{2} m^2 a^2
-  \frac{1}{4!} \frac{m^2 }{\fPQ^2} a^4 
+  \frac{1}{6!} \frac{m^2 }{\fPQ^4} a^6
~+ ...
\label{eqn3}
\eeq
and  the axion mass is
\beq
m \simeq \frac{m_\pi f_\pi}{\fPQ}  \frac{\sqrt{m_um_d}}{m_u+m_d}
 \simeq  10^{-4} ~{\rm eV}~ 
\frac{6\times 10^{10} \, \rm GeV}{\fPQ} \, .
\label{mf}
\eeq
In this paper, we take $m \simeq 10^{-4}$ eV,
because in the scenario
where the Peccei-Quinn phase transition is
after inflation,  the numerical simulations  of
Kawasaki, Saikawa and Sekiguchi~\cite{KSS} (hereafter KSS)
suggest that this gives the correct dark matter abundance.
$\fPQ$ is the breaking scale of the Peccei-Quinn symmetry,
here taken to be fixed in terms of the axion mass by eqn
(\ref{mf}). The potential (\ref{eqn3}) is therefore
a one-parameter potential determined by $m$.

In the non-relativistic limit, the real axion field
can be written in terms of a complex field $\phi$ \cite{NambuSasaki}
\beq
a = \frac{1}{\sqrt{2m}}(\phi e^{-imt} + \phi^* e^{imt})
=  \frac{1}{\sqrt{2m}}(\eta e^{-i(S+mt)} + \eta e^{i(S+mt)}) \,,
\label{NRa}
\eeq
with $\eta$ and $S$ real. 
It is intuitive that a real field becomes complex
in the non-relativistic  limit,
because particle number is conserved. The potential
for the non-relativistic field is
\beq 
V(\phi) =  \frac{m}{2} \phi^* \phi  -    \frac{g}{2} |\phi|^4 
+...
\,,\qquad
\frac{g}{2} =  \frac{1}{16\fPQ^2} 
\label{Veta}
\eeq
(obtained by dropping the terms that oscillate as $e^{\pm iNmt}$, 
on the assumption that they average to zero), and the field $\phi$
satisfies a Schr\"odinger-type equation
\beq
i \dot{\phi}  =  -
\frac{\nabla^2}{2m} \phi  
 - |g| (\phi^\dagger \phi)
\phi  
+  m V_N \phi  ~~~,~~~
~~~~~~~~{\rm GP~ equation} 
\label{GP}
\eeq
(obtained by neglecting $\partial_t^2$ and $(\partial_t)^2$ terms in
the Klein-Gordon equation for $a$, see appendix for a derivation).
Note that in ``natural units'' $\phi$ and $g$ have mass dimension 3/2
and $-2$, respectively, while the Newtonian potential $V_N$ is
dimensionless\footnote{Recall that eqn~(5) is a classical field
  equation, so it contains no $\hbar$, despite its formal similiarity to
  the Schr\"odinger equation.  Setting $c=1$ (time in units of
  distance $d$), the dimensions are $[\phi]=\sqrt{E}/d$, $[g]= d/E$
  and $[m] =1/d$, where $E$ is energy or mass.  In particular, the
  parameter $m$ of the classical field is an inverse length, and
  $\hbar$ is required to relate it to the mass of quanta of the field
  \cite{DavidsonElmer, h-expansion}.}.  Equation (\ref{GP}) is
referred to as the Gross-Pitaevski (GP) equation, and is widely used,
from describing Bose Einstein condensation of cold atoms to galaxy
halos made of $m\sim 10^{-20}$ eV bosons. A useful review about this
equation is \cite{DGPS}.

The dynamics of the axion field coupled to gravity can 
be obtained from $T^{\mu\nu}_{~~;\nu}=0$, or from the
Klein Gordon equation. The axion $\Tmn$ is parametrised
by the axion energy density $\rho$ and fluid three-velocity $\vec{v}$,
which are more intuitive variables for Large-Scale-Structure
(LSS) than the field. The transformation between
these two parametrisations is discussed in the
works of Chavanis\cite{Chavanis} and Rindler-Daller
and Shapiro (RDS) \cite{RDS}.
The stress-energy tensor
 for the  non-relativistic axion field, in
cartesian coordinates for 
flat space-time (Newtonian gravity can be added later by hand),
 is
\bea
T_{00}& = & \rho = m\eta^2 + ...\nonumber \\
T_{0i} &=&\eta^2 \partial_i S   + ... = -\rho v^i +...\nonumber\\
 T_{ij}& =&\frac{1}{2m} {\Big (}  
 2\partial_i \eta  \partial_j \eta
+  2\eta^2  \partial_i S  \partial_j S 
+ \delta_{ij}[   -\nabla   \eta
\nabla   \eta  +   2m \eta^2 \dot{S}- \eta^2 \nabla  S \nabla  S  
+ m |g| \eta^4 ]  {\Big )} 
\label{Tij} \\
 &=&\frac{\partial_i \rho  \partial_j \rho}{4m^2 \rho} 
+  \rho v_i v_j 
 -\delta_{ij} \left[
 \frac{\nabla^2  \rho }{4m^2 }
- \frac{|g|}{2 m^2} \rho^2 \right] 
\eea
 where  the equations of motion
were used to simplify the brackets  between
the first and   second lines  for $ T_{ij}$, and we defined
${v}^j = -\partial_j S/m$.


Two equations are obtained from  $T^{\mu\nu}_{~~;\nu}=0$: 
\begin{align}
\partial_t \rho = &  - \nabla \cdot \rho \vec{v} & {\rm continuity}  
\label{continuity} \\
\rho \partial_t \vec{v}
+ \rho \vec{v} \cdot \nabla \vec{v}  =&  
\rho \nabla
{\Big (}
\frac{ \nabla^2 \sqrt{ \rho}}{2m^2\sqrt{ \rho}}
+ |g| \frac{\rho}{m^2}  
-   V_N{\Big )}  & {\rm Euler}  \label{euler}
\end{align}
which can also be obtained from the  real and imaginary parts
of the complex
GP equation (\ref{GP}) by using
$\phi = \sqrt{\frac{\rho}{m}}e^{-iS}$ and
${v}^j = -\partial_j S/m$, and taking the
divergence of the real equation. 
The gravitational potential $V_N$ is obtained from
the Poisson Equation
\beq\label{poisson}
\nabla_x^2 V_N(x-x') = 4\pi G_N \rho(x') \,,
\eeq
which  outside a
spherical mass distribution,  has the familiar solution $V_N(r) = - G_N M(r)/r$
with $M(r)$ being the mass inside the radius $r$.

From the Euler equation, one can already see that a stationary solution
(neglecting rotation, so setting the left side of Euler to zero)
should balance the outwards  gradient pressure represented by
the first term, against the inwards gravitational and
self-interaction pressures. The self-interaction pressure
is inwards because the density decreases with $r$, so 
$\partial_r\rho <0$.
An estimate for the mass $M$ and radius $R$  of an axion drop can be obtained
by replacing $\nabla \to 1/R$, $\rho \to M/R^3$ 
on the right-hand-side of the Euler equation, and solving the resulting
quadratic equation for  $R$:
\beq
 {\Big (}
\frac{1}{2m^2 R^2}
- |g| \frac{M}{m^2R^3}  
- G_N \frac{M}{R} {\Big )}\simeq 0 
~~~\Rightarrow ~~~
R  \sim \frac{\mpl^2}{4m^2M}
~~,~~
M \lsim \frac{ \mpl \fPQ}{m} \,.
\label{estimate}
\eeq
This exhibits an upper bound on the drop mass, as found by Chavanis,
as well as the usual ``virial'' relation between the radius
and mass of an object supported against gravity by pressure.
Below we will use the virial theorem to obtain a more
reliable equation, but this already indicates the
parametric dependence of the mass and radius of the drop.
Notice that the  maximum drop mass $\propto 1/m^2$, so
lighter axions can form larger drops.
Solving for $R$  at the maximum mass 
$ M \sim 10^{-14} M_\odot$,
 gives  $R \sim 100$ { km}.

More sophisticated estimates are obtained,
for instance, by  Chavanis or RDS, 
by guessing a functional form for
$\rho(\vec{r})$, and minimising the energy functional that
gives the GP equation. RDS consider bosons of
variable mass and  positive  self-interaction (which provide
an outward pressure, helpful in obtaining  cores for galactic
halos), and find rotating field configurations which could
be galactic halos for bosons of mass $m \sim 10^{-20}$ eV. 
Chavanis considers variable masses and self-interactions
of either sign,  and looks for
non-rotating solutions. For a QCD axion (negative
self-interaction), and  a non-rotating drop, Chavanis found
a maximum mass of  $10 \mpl \fPQ / m$,
by minimising an energy functional with an exponential ansatz
 for the radial profile of the density.

Estimates similiar to eqn~(\ref{estimate}) can be obtained
using the virial theorem. 
 It says, for a stationary,
spherically symmetric space-time \cite{Wald}
\beq
\frac{3}{2} \int P r^2 dr d \Omega  = \int G_N \frac{\rho M(r)}{r} r^2 dr d \Omega
\label{Wald}
\eeq
where $3P = \sum_i T^{ii}$. This can be written in the form used
by Chavanis and RDS 
\beq
 E_{grav} + 2E_{cin}  + 3E_{si} = 0 \,,
\label{virial}
\eeq
 where
\beq
E_{grav} = \int dV \frac{\rho}{2}V_N \,,~~~
E_{si} = g \int dV \frac{\rho^2}{2m^2} \,,~~~
E_{cin} = \frac{1}{2}\int dV \left[\frac{(\nabla \rho )^2}{4 \rho m^2}
+  \rho |\vec{v}|^2\right] \,.
\eeq


We  assume in this paper, following
RDS,  that a rotating
axion drop should also satisfy   
the virial condition eqn~(\ref{virial}).  
We consider rotating drops of axion field, 
and estimate whether the rotation could allow them 
to be significantly more massive than the estimate of 
eqn~(\ref{estimate}). 
To  obtain an ansatz for the rotating drop,   
notice the ressemblance between 
the (non-linear)  GP equation and the  (linear)
Schrodinger equation for the Hydrogen
atom, the latter having  well-known solutions 
in terms of spherical harmonics.  
 It is convenient
to  start from the axion field, rather than
$\rho$ and $\vec{v}$, because the  phase of
the field  should be continuous, and this condition
is less simple to impose on $\rho$ and $\vec{v}$. 
  Following   Tkachev~\cite{IgorBoseStar}, we 
 suppose the rotating axion drop has the
functional form in radial coordinates
\beq
\phi(r,\theta,\varphi) =\sqrt{\frac{\rho_c}{m}}
F(r) Y^l_l(\theta,\varphi) \,,~~~
 Y^l_l = c_l\sin ^l \theta e^{il\varphi} \,,~~~
c_l =  \frac{(-1)^l}{2^l l!}\sqrt{\frac{(2l+1)!}{4\pi}}
\label{Yl}
\eeq
where $c_l$ is taken 
such that $\int |Y^l_l|^2 d\Omega = 1$, which ensures
that the total mass of the drop is independent of $l$
and remains $\sim \rho_c r_c^3$.
For simplicity, the radial function $F(r)$  is
taken   to give a density of 
top-hat form: 
\beq
\rho(r,\theta) =
|c_l|^2 \rho_c 
\theta(r_c -r)
 \sin ^{2l} \theta \,,
\label{rhol}
\eeq
in which case the total mass of the drop is $M = \rho_c  r_c^3/3$. 

Various comments can be made.
\ben
\item The  radial density profiles 
$F(r)$  contain two parameters:
 a central density $\rho_c$  (integrated
over angles), and $r_c$ which is  some measure of the size
of the drop. In addition to
the top-hat,  we tried an 
 ``isothermal-sphere-squared'' profile,
$F(r) = r_c^2/(r^2 + r_c^2)$, 
because it
approaches being a solution of the static  Euler equation and 
 has a finite  volume integral.  These results
are not given, because  they  only differ
from the top-hat profile in irrelevant 
numerical factors (despite that the 
integration is more involved).

\item A solution of the Schr\"odinger equation is 
usually expanded  on  the set of  $\{ Y^l_n\}$.
We select one $Y^l_l$ for simplicity; the equations
of motion are non-linear, so  this allows to avoid
products $Y^{l}_{n} Y^{* l'}_{n'}$.
In addition,
the parametrisation
$\phi  = \sqrt{\frac{\rho}{m}}e^{-iS} \propto Y^l_l$,
$\vec{v} = -\nabla S/m$, relates $l$ 
to the fluid velocity $v_r = 0, v_\theta = 0, v_\varphi = l/(mr \sin
\theta)$. Note that in the case of a single $Y^l_n$ the choice $Y^l_l$
with $n=l$ corresponds to chosing the $z$-axis of the coordinate
system along the angular momentum vector. 


\item 
Asteroids in our  solar
system  can have masses and radii  comparable to the non-rotating axion drops, and  tend to  have rotation periods $\sim 6$ hours. However,
their formation history differs from that of axion drops, so
it is unclear whether this is a relevant  analogy. 
The equatorial rotation frequency of a drop described by
eqn~(\ref{Yl}), evaluated at 
the radius $r_c$,   would be
$\omega \simeq {l}/{(r_c^2 m)} \simeq 6l$/day,
which  suggests that low $l$ values are realistic.

\item With the ansatz of 
eqn~(\ref{Yl}),
the parameter $l$ describes two  distinct
physical aspects of the drop: 
its rotation, and also   its  flattening 
into a disk.   However, we allow  this degeneracy, because
we  only consider $l$ values of order a few, due to the
previous point.

\een


To obtain the gravitational energy
which enters the virial condition (\ref{virial}), the
 potential $V_N(r,\theta)$ is required. Expanding  the
density  on spherical harmonics,
$\rho (r,\theta) =  \rho(r) \sum_{k \leq 2l} \eta_k  Y^k_0 (\theta)$
with  $\eta_k = \int d\Omega \sin^{2l} \theta Y^k_0 $, the
potential is~\cite{BT}
\bea
V_N(r,\theta) = -4 \pi G_N \sum_k \frac{\eta_k  Y^k_0(\theta)}{2k+1}
\left( \frac{1}{r^{k+1}} \int _0 ^r da a^{k+2} \rho(a) 
+ r^k \int _r ^\infty \frac{da }{a^{k-1}} \rho(a) \right)
\equiv  -4 \pi G_N \sum_k   V_k(r) Y^k_0(\theta)
\eea
which illustrates the interest of the top-hat density profile. This gives
\bea
E_{grav} & =& -2 \pi G_N \sum_k \int_0^\infty r^2 \eta_k  \rho(r) V_k(r) dr~~~\nonumber \\
&= & - \frac{4\pi G_N}{15} \rho_c^2 r_c^5 |c_l|^4
\left\{
|\eta_0|^2
+  \frac{3|\eta_2|^2}{5} +
 \sum_{4 \leq k\leq2l} \frac{3|\eta_k|^2}{(k+3)(2k+1)} \right\} \nonumber\\
 &\gappeq&
- \frac{4\pi G_N}{15} \rho_c^2 r_c^5 \frac{|c_l|^4 }{|c_{2l}|^2}
\longrightarrow  - \frac{3 G_N M^2}{5 r_c}  \sqrt{1+l}~~~~
\eea
where, in the last approximation,
the curly brackets were taken $\leq \sum_k |\eta_k|^2= 1/|c_{2l}|^2$,
and after the arrow 
approximates\footnote{
For $l = \{0,1,2,3,...\}$,   
$\frac{|c_l|^4}{|c_{2l}|^2} = \{ 1, \frac{5}{6}, \frac{5}{7}, 1.63,...\}
\times \frac{1}{4\pi}$. 
For large $l$, the  Stirling approximation 
for large-$n$ factorials,
 $n! \simeq \sqrt{2\pi n} \left( \frac{n}{e} \right)^n$, gives
$\frac{|c_l|^4}{|c_{2l}|^2}
= \sqrt{\frac{2l}{\pi}} \frac{1}{4\pi}$.} 
$\frac{|c_l|^4}{|c_{2l}|^2} \simeq \sqrt{l+1}/4\pi$.
The kinetic and
self-interaction energies are
\bea
3 E_{si} &=& -\frac{M^2}{ r^3_c}  \frac{9}{ 16  \fPQ^2 m^2 }
\frac{|c_l|^4 }{|c_{2l}|^2}
\longrightarrow -\frac{M^2}{ r^3_c}  \frac{9}{ 64 \pi  \fPQ^2 m^2 } \sqrt{1+l}
\label{ESI}\\
2 E_{cin} &=& \frac{3M}{8m^2 r_c ^2}[1+ 4 l(l+1)] \,,
\label{Ecin}
\eea
using $(\partial \rho/\partial r)^2 \simeq \rho_c^2 |c_l|^4 \sin^{4l}\theta \, \delta(r-r_c)/r$.

We see that the gravitational and self-interaction energies grow
as $\sqrt{l+1}$. This is because the drop flattens into a disk
for large $l$ (as a result of our ansatz (\ref{rhol})),
so the density is larger to stay at constant mass. The kinetic energy
grows quadratically with $l$, both because
the drop has angular momentum, and because  of the
gradient in $\theta$. Solving the Virial equation for $r_c$ gives
 \beq
M r_c  \simeq \frac{5\mpl^2}{8m^2} \,
\frac{1 + 4l(l+1)}{\sqrt{l+1}}
\left[1 \pm \sqrt{ 1 - \frac{3 (l+1)m^2M^2}{5 \pi [1 + 4l(l+1)]^2 \fPQ^2 \mpl^2 }}\right]
~~\Rightarrow~~
M \lsim
\sqrt{\frac{5\pi}{3}}
\frac{\mpl \fPQ}{m}\frac{1 + 4l(l+1)}{\sqrt{l+1}}
\label{estimate2}
\eeq
which, for $l=0$, reproduces the  parametric dependence
of the estimates obtained from the Euler equation  in 
eqn~(\ref{estimate}). 
However, the constants are unreliable, because Chavanis
obtained a maximum mass of $\sim 10^{-13}~ M_\odot$  with
an exponential ansatz for the radial density profile of
a spherically symmetric solution.  So we suppose the upper
bound is
\beq
M\lsim \frac{1 + 4l(l+1)}{\sqrt{l+1}} \times  10^{-13}~ M_\odot \,.
\label{Mdrop}
\eeq
For $l \sim$ few, which corresponds to the rotation
rates of asteroids, the maximum mass of the drop can
grow by an order of magnitude or so. This behaviour
is confirmed by the numerical calculations presented
in the next section.

The upper bound on the size of the drop arises
because the  self-interaction energy of the QCD axion
is negative (equivalently, it exerts an
inwards force, like gravity): if the 
mass $M$ inside a volume $\sim R^3$
is to high, the gradients cannot compensate 
the self-interaction energy.  However, the QCD
axion has a cosine potential, see
eqn~(\ref{eqn3}), and not the unbounded-below
$-|g| |\phi|^4$ potential used to compute $E_{si}$.
So one can wonder whether the maximum mass is
an artifact of expanding the potential.
It seems no: a more correct version
of the potential, in the non-relativistic approximation where
the mass term does not appear, 
would be $m^2 \fPQ^2[1 - \cos(a/\fPQ)] - m^2 a^2/2$.
This is also always negative,
and well-approximated by $-m^2 a^4/(4! \fPQ^4)$ 
at the maximal drop mass
(which corresponds to $a/\fPQ \sim  \fPQ/\mpl \ll 1$).
So  drops beyond the maximal mass
would collapse\footnote{One could wonder if
 there are  smaller, heavier drops, whose
radius  could be estimated by balancing the 
inward self-interaction
pressure from $m^2 \fPQ^2[1 - \cos(a/\fPQ)] - m^2 a^2/2$,
against the outwards gradients.
However, the radius where this self-interaction
energy balances the gradient energy can be estimated
to be $\sim 1/m$.} or fragment.
Ref.~\cite{BB} shows by numerical calculation that the 6th order term
in the potential has a small impact on the density profile.


\section{Numerical solution of the GPP system}
\label{sec:num}

In this section we are solving the coupled Gross-Pitaevski
[eqn~\eqref{GP}] and Poisson [eqn~\eqref{poisson}] equations
numerically. We proceed in analogy to the standard treatment of the hydrogen
atom.  We make an ansatz for the field in terms of a radial wave
function $R(r)$ and spherical harmonics:
\begin{align}\label{eq:psi-ansatz}
  \phi = R(r) Y^l_m(\vartheta, \varphi) e^{-iEt} \,,\qquad M = m \int_0^\infty dr \, r^2 R^2(r) \,,
\end{align}
(so $R(r) = \sqrt{\frac{\rho_c}{m}} F(r)$ in eqn~(\ref{Yl})).
In general $\phi$ may consist of a superposition of
various $Y^l_m$.  As before, we assume here for simplicity that the
field consists just of a single $Y^l_m$ with given $lm$. As usual,
thanks to the factor $e^{-iEt}$ we obtain a time-independent
Schr\"odinger equation:
\begin{align}\label{eq:schr1}
  ER = \left\{ -\frac{1}{2m} \left[ \Delta_r - \frac{l(l+1)}{r^2} \right]  +
  m V_N - |g| R^2|Y^l_m|^2 \right\} R \,,
\end{align}
where $\Delta_r$ is the radial part of the Laplace operator in polar coordinates.
%
Note that this equation still depends on $\vartheta$ and $\varphi$ via
the potential $V_N$ and the self-interaction term. Therefore, we take
the angular average of eqn~\eqref{eq:schr1}.  We define the angular
averaged gravitational potential and use the normalization of the
spherical harmonics:
\begin{align}\label{eq:aver}
  \overline V_N(r) \equiv \frac{1}{4\pi} \int d\Omega \, V_N(r,\vartheta,\varphi) \,,\qquad
  \int d\Omega |Y^l_m|^2 =  1 \,.
\end{align}
Then we obtain for the angular averaged GP and Poisson
equations: 
\begin{align}
  \Delta_r R &=
  \left[ 2m(m \overline V_N - E) + \frac{l(l+1)}{r^2} - |g| \frac{m}{2\pi}R^2 \right]R \\
  \Delta_r {\overline V_N} &= G_N m R^2 \label{eq:Paver}\,,
\end{align}
which now are just two coupled second order differential equations for the two functions
$R(r)$ and $\overline V_N(r)$.

We can make this system dimensionless and absorb the coupling
constants $g$ and $G_N$, the axion mass $m$, and the constant
$E$ by appropriate rescaling:
\begin{align}
  \Delta_{\tilde r}{\tilde R} &=
  \left[ \tilde V_N + \frac{l(l+1)}{{\tilde r}^2} - {\tilde R}^2 \right] \tilde R
   \label{eq:GPP1}\\
   \Delta_{\tilde r} {\tilde V_N}&= {\tilde R}^2 \,,
   \label{eq:GPP2}
\end{align}
with
\begin{align}
  r = \frac{1}{m} \sqrt{\frac{|g|}{4\pi G_N}} \, \tilde r \,,\qquad
  R = 2\pi \frac{\sqrt{2G_N m}}{|g|} \tilde R \,,\qquad
  \overline V_N - \frac{E}{m} = \frac{2\pi G_N}{|g|} \tilde V_N \,,
\end{align}
where the tilde quantities are dimensionless.
Using $\mpl = 1/\sqrt{G_N} \approx 1.2\times 10^{19}$~GeV and the QCD expression for $g$
given in eqn~\eqref{Veta} we can write those relations as
\begin{align}
  r &= \frac{1}{4\sqrt{2\pi}} \frac{\mpl}{m \fPQ} \tilde r
  \approx 4\times 10^4\,{\rm m}\, \tilde r\,, \qquad
  R = 16\sqrt{2}\pi \frac{m^{1/2} f_{\rm PQ}^{2}}{\mpl} \tilde R \,,\qquad
  \overline V_N - \frac{E}{m} = 16\pi \frac{f_{\rm PQ}^2}{\mpl^2} \tilde V_N \,.
  \label{eq:tilde-R}
\end{align}
The total mass of the axion drop is given as
\begin{align} \label{eq:tilde-M}
  M = 2 \sqrt{2\pi} \frac{\mpl f_{\rm PQ}}{m} \, \tilde M
  \sim 10^{-13} M_\odot \left( \frac{f_{\rm PQ}}{10^{11} \, \rm GeV} \right)^2 \, \tilde M 
\qquad
\text{with}
\qquad
  \tilde M = \int_0^\infty d\tilde r \, {\tilde r}^2  {\tilde R}^2 \,.
\end{align}
Assuming that tilde quantities are of order one,
eqs~\eqref{eq:tilde-R} and \eqref{eq:tilde-M} set the scales for the
typical dimensions of the axion drop, in agreement with the estimates
of the previous section. This confirms that one obtains the
same physics by studying either the equations of motion
for the field,  or Einsteins equations for the stress-energy
tensor.  We can also estimate the typical density of the drop as
\begin{equation}
  \rho(r) = m R^2(r) \approx 3 \, {\rm g\, cm^{-3}} \left( \frac{f_{\rm PQ}}{10^{11} \, \rm GeV} \right)^2 \, {\tilde R}^2(r) \,, \label{eq:density}
\end{equation}
which is comparable to the average density of the Earth for this choice of $\fPQ$.


Now the task is to numerically solve the GPP system \eqref{eq:GPP1},
\eqref{eq:GPP2}.  These are 2 second order differential equations for
the functions $\tilde R(\tilde r)$ and $\tilde V_N(\tilde r)$. We need
to specify 4 initial conditions: $\tilde R(0), \tilde R'(0), \tilde
V_N(0), \tilde V_N'(0)$, with prime denoting derivative with respect
to $\tilde r$.

For a given $R(r)$  we can integrate the
Poisson equation \eqref{eq:GPP2} and obtain a solution for the potential:
\begin{equation}
  \tilde V_N(\tilde r) = - \frac{1}{\tilde r} \int_0^{\tilde r} dx \, x^2 {\tilde R}^2(x) 
   - \int_{\tilde r}^\infty dx \, x {\tilde R}^2(x) + \tilde E \,.
\end{equation}
Here $\tilde E$ is an integration constant. Another integration constant has been chosen such
that we obtain the following limiting expressions:
\begin{align}
  \tilde V_N(0) &=  
   - \int_0^\infty dx \, x {\tilde R}^2(x) + \tilde E \,,
   \qquad
   \tilde V_N(\tilde r \to \infty) = - \frac{\tilde M}{\tilde r} + \tilde E  \to \tilde E\,.  
\end{align}
This choice of integration constants implies that
$\tilde V_N(\tilde r)$ is finite at $\tilde r = 0$ and
${\tilde V_N}'(0) = 0$. At distances far away from
the mass distribution we obtain the Newtonian result
for the potential in physical units, $V_N(r) = - G_N
M / r$, when we identify
\begin{equation}\label{eq:tilde-E}
  \tilde E = - \frac{|g|}{2\pi G_N} \frac{E}{m} 
           = - \frac{1}{16\pi} \frac{\mpl^2}{f_{\rm PQ}^2}  \frac{E}{m} \,.
\end{equation}
Considering the GP equation at large radii and requiring that $R(r)$
decreases exponentially we find that $\tilde E$ should be positive,
i.e., $E < 0$. In the familiar Schr\"odinger equation for
the hydrogen atom, $E$ would
be the binding energy of the electron. We find this analogy
useful, despite that, in our classical field
theory, the energy  is a volume integral
and $E$ has units of frequency.
 As shown below, for the solutions
we have $\tilde E \lesssim 1$.  Hence it follows
\begin{equation}\label{eq:E}
-\frac{E}{m} \lesssim 16\pi \frac{f_{\rm PQ}^2}{\mpl^2}  
\approx 3.5\times 10^{-15}\left( \frac{f_{\rm PQ}}{10^{11} \, \rm GeV} \right)^2 \,  
\end{equation}
and $|E| \ll m$, which justifies the nonrelativistic treatment.

\begin{figure}[t]
  \centering
  \includegraphics[width=0.8\textwidth]{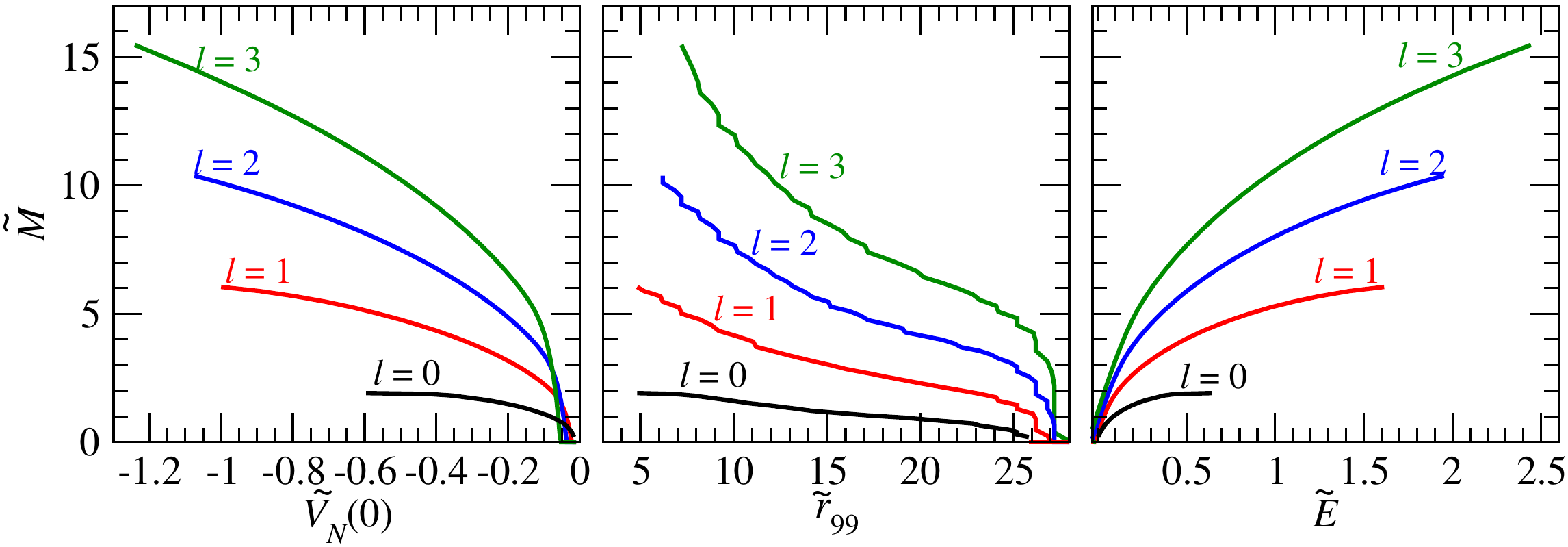}
  \caption{A set of solutions of the GPP system for $l=0,1,2,3$. The 3 panels show the total
    mass as a function of the initial values for $\tilde V_N$ (left),
    as a function of the radius containing 99\% of the mass (middle),
    and $\tilde E$ (right). The initial value
    $\tilde R(0)$ has been fixed at $10^{-2}$ for $l=1$ and at $10^{-4}$
    for $l=2,3$, while for $l=0$ it has a finite value.  The relation of the
    dimensionless quantities shown in
    the plots to physical quantities is given in
    Eqs~\eqref{eq:tilde-R}, \eqref{eq:tilde-M},
    \eqref{eq:tilde-E}.
    \label{fig:mass-l}}
\end{figure}

\begin{figure}
  \centering
  \includegraphics[width=0.41\textwidth]{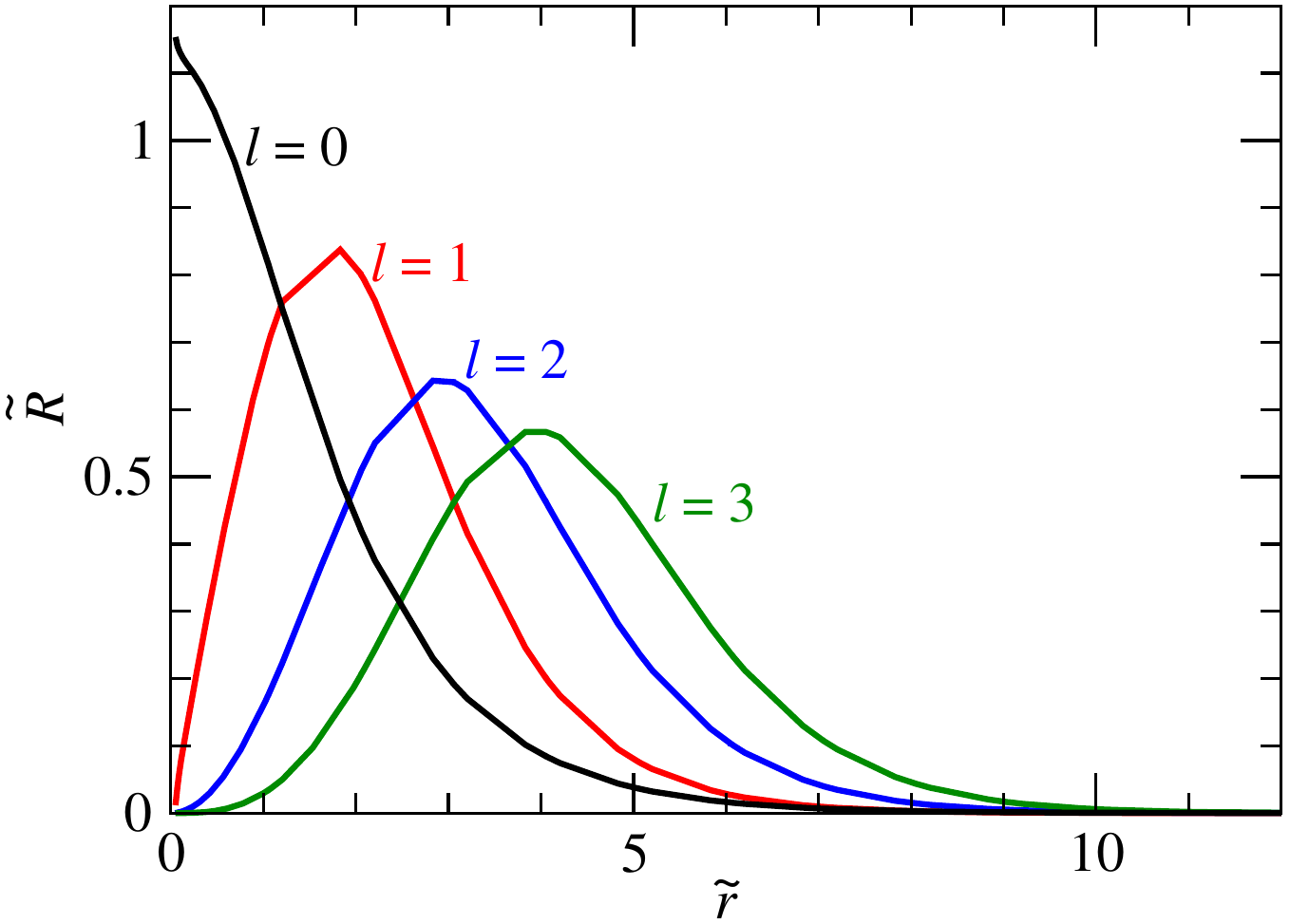} \qquad
  \includegraphics[width=0.4\textwidth]{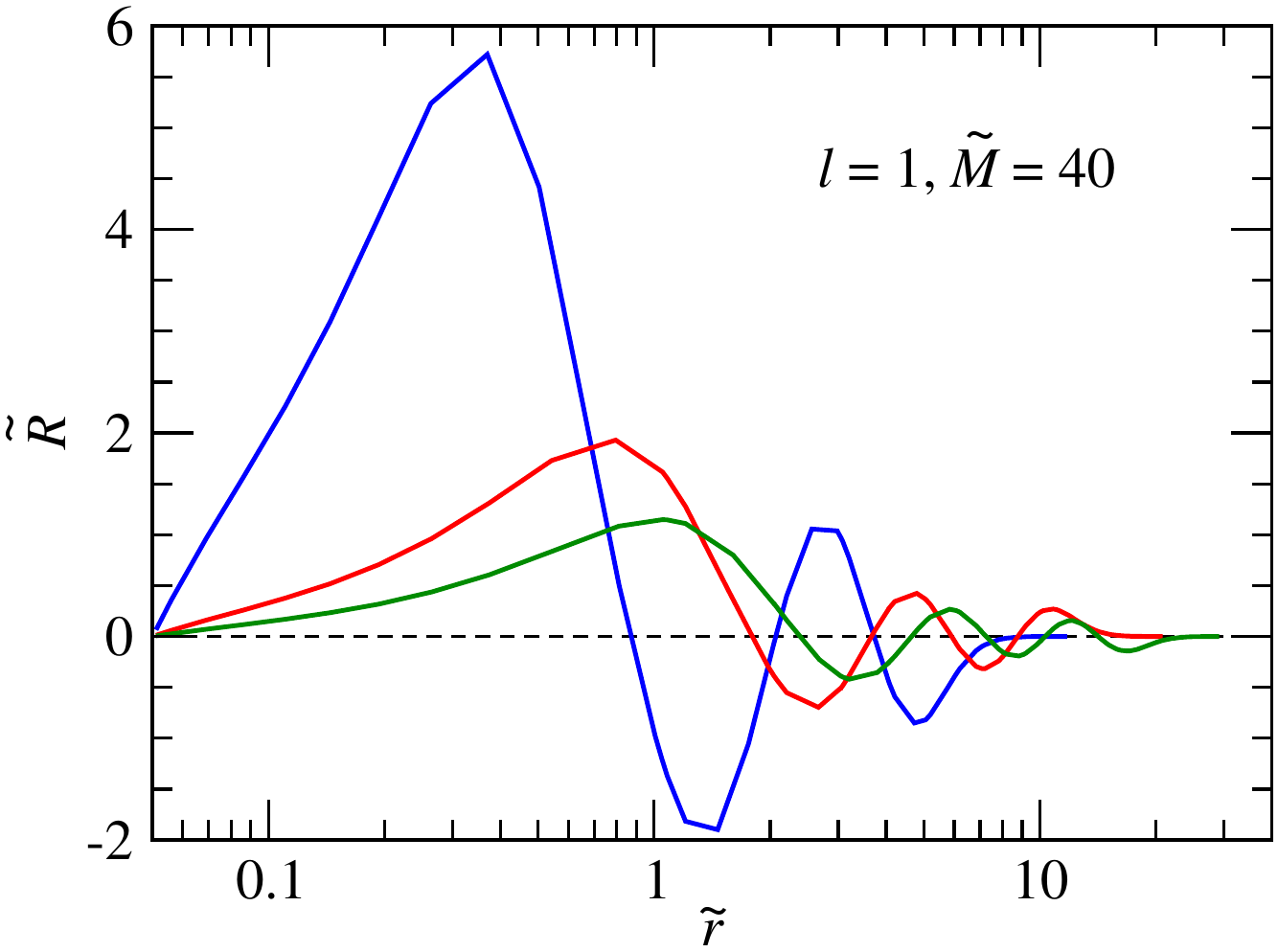}
  \caption{Left: The radial wave function $\tilde R(\tilde r)$ for the
    solutions for $l=0,1,2,3$
    corresponding in each case to the maximum mass shown in figure~\ref{fig:mass-l}.
    Right: Radial wave function for solutions for $l=1$ with 3, 4, and 5 nodes, where
    all three examples correspond to a total mass $\tilde M \approx 40$.
\label{fig:profile-l1}}
\end{figure}

Our procedure is now as follows: We set $\tilde V_N'(0) = 0$ and pick
some values for $\tilde R(0)$ and $\tilde V_N(0)$. Then we search for
$\tilde R'(0)$ by a shooting method, such that for the solution $\tilde R$, ${\tilde R}'$
and ${\tilde V_N}'$ go to zero at large $\tilde r$.  If a solution is
found we can then calculate the corresponding total mass $\tilde M$,
the radius containing 99\% of the mass, $\tilde r_{99}$, and $\tilde E$.
In Fig.~\ref{fig:mass-l} we show the behaviour of the total mass of
the drop, $\tilde M$, for the cases $l=0,1,2,3$. Beyond the point
where the curves stop we could not find any solution 
with physical boundary conditions,
indicating the existence of a maximal possible mass.
When we try to extend the curves beyond the
maximal mass, the solutions diverge at large radii. In
Fig.~\ref{fig:profile-l1} (left panel) we show the radial profiles for
the maximal mass for each $l$ value. For those solutions we required
zero nodes of $\tilde R(r)$.


For $l>0$ physically interesting solutions have vanishing wave
function at the centre, since the angular momentum term in the
GP equation diverges as $r \to 0$ for finite $R(0)$. For
the numerics, we fix $\tilde R$ at 0.01 for $l=1$ and at $10^{-4}$ for
$l=2,3$ at radius $\tilde r = 0.05$ and then we scan different values
of $\tilde V_N(0)$.  For each value we search for the derivative
${\tilde R}'(0)$ in order to obtain a physically interesting solution,
converging at large radii. For $l=0$, $\tilde R(0)$ is non-zero and we
have to scan also over this parameter, in addition to $\tilde V_N(0)$
and ${\tilde R}'(0)$.

Our results for the $l=0$ case are in qualitative agreement with
figs.~4 and 5 from \cite{Eby:2015hsq}. However, sofar we could not
find solutions with smaller $\tilde r_{99}$ as the one corresponding
to the maximal mass (cf.\ fig.~4 of \cite{Eby:2015hsq}) or
equivalently larger $\tilde R(0)$ (cf.\ fig.~1
of \cite{Barranco:2012ur}).  Note that to search for $l=0$ solutions
we have restricted the derivative ${\tilde R}'(0)$ to be small (the
condition we impose is $|{\tilde R}'(0)| < {\tilde R}(0)$).  If we
allow for large derivatives at small radii solutions are found for
different combinations of $\tilde R(0)$ and $\tilde V_N(0)$. Those
allow also for slightly larger total masses, although qualitatively
the behaviour is similar to the shown solutions.

We find that the solutions for $l > 0$ are more massive than in the
$l=0$ case by about a factor 3 for $l=1$, a factor 5 for $l=2$, and a
factor 8 for $l=3$, but the size increases only slightly, from $\tilde
r_{99} \approx 5$ to 7. Restricting to solutions with zero nodes we
did not find any solution for $l \ge 4$.

\begin{figure}
  \centering
  \includegraphics[width=0.7\textwidth]{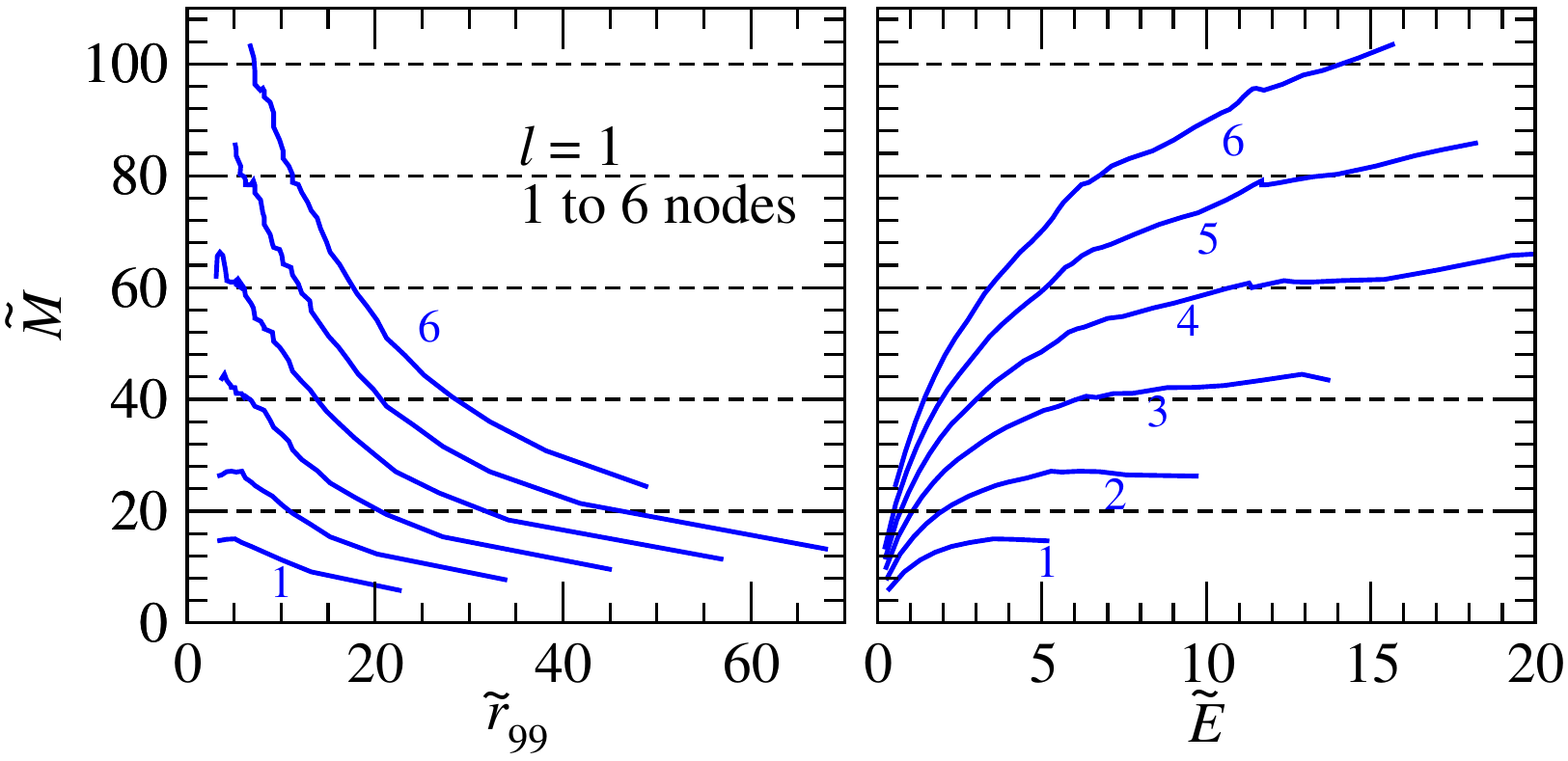}
  \caption{Solutions for $l=1$ with 1 to 6 nodes of the radial wave
    function (curves from bottum up have increasing number of nodes). We show
    the total mass $\tilde M$ as a function of $\tilde r_{99}$ (left)
    and $\tilde E$ (right).
    \label{fig:multi-node}}
\end{figure}

In Fig.~\ref{fig:multi-node} we show solutions with 1 to 6 nodes for
the case with $l=1$. We observe that significantly larger masses can
be obtained for multi-node solutions. Fig.~\ref{fig:profile-l1} (right
panel) shows three examples for the wave function with $l=1$ and 3, 4,
and 5 nodes. For all three examples the total mass is approximately
$\tilde M \approx 40$. We have also searched for multi-node solutions
with higher values of $l$. As in the zero-node case, we have not found
physical solutions for $l \ge 4$.

The results of our numerical study of the GPP system can be summarized
as follows.
\begin{enumerate}
\item 
We recover the well known result from the literature that
for the ``ground state'' with zero nodes and $l=0$ there is a maximal
possible mass, let's call this maximal mass of the ground state
$\tilde M_0^\text{max}$.

\item
For a given mass, solutions with lower angular momentum $l$ and less
nodes have larger  $\tilde E$ and hence are
favoured.  However, for masses $\tilde M > \tilde
M_0^\text{max}$, there seems to be a minimal $l$ and (for given $l$) a
minimal number of nodes, for which solutions exist.  Hence, if we look
for a solution for a fixed mass with $\tilde M > \tilde
M_0^\text{max}$, the solution with the maximum  $\tilde E$ is obtained
at $l > 0$ and/or with a non-zero number of nodes.

\item
We have not found solutions with $l\ge 4$;  this may be
an artifact of our single $Y^l_m$ approximation, or may
indicate an upper bound on the angular momentum of the drop.
It  deserves further study.
\end{enumerate}

In this paper we have not considered the question of stability of
solutions against small perturbations.  Ref.~\cite{Lee:1988av} has
found that for the non-interacting and non-rotating case, only the
ground-state with zero-nodes is stable, whereas solutions with nodes
are unstable with respect to oscillations, see also
\cite{Jetzer:1991jr}. We leave the stability analysis of multi-$l$ and
multi-node solutions for the QCD axion case for future
investigations. This question needs to be addressed as well within
the dynamical framework of the formation of the axion drops.


\section{From the QCD phase transition till today}
\label{secB}

\subsection{Estimating the  minicluster mass}

We suppose that the Peccei-Quinn phase transition  occurs
after inflation, leaving  a massless axion field which is random
from one horizon-volume to the next, a network of
cosmic strings, and no domain walls. 
Shortly before the QCD phase transition,
the axion mass  is expected to turn on,  generating
the potential (\ref{eqn3}).
This leads to two axion contributions to dark matter:
\ben
\item It causes the network of cosmic strings to decay. This
process is simulated numerically~\cite{HKSS,DS89,BS94}; the recent results
of KSS \cite{KSS}
give two comparable  contributions to axion dark matter.
From the Peccei Quinn phase transition until the axion mass turns on,
the string network radiates axions with $p\sim H$.
Then   domain walls form
between the strings, separating regions of $\pm a$,
and subsequently this wall-string network decays away to
axions  with $p\sim m(T) \sim H$.

\item The axion field, random in each horizon-volume,
is likely to be misaligned with respect to the minimum
of the potential, so will roll down and oscillate.
The initial magnitude of the field, averaged over many horizons,
is $\pi\fPQ/\sqrt{3}$. 
The oscillations redshift like cold dark matter~\cite{DineFischler,PWW,BHK}. 
\een

Combining these contributions to axion dark matter, KSS \cite{KSS}
obtained an appropriate relic density for $m \sim 10^{-4}$~eV.
Only $\sim 25\%$ of this CDM is due to the misalignment
field, the remainder is in the  incoherent distribution
of axion modes produced by strings. 

In this paper, we are interested in the misalignment
axion field, which can have two types of density fluctuations. 
It inherits the  large-scale adiabatic density
fluctuations present in radiation, which will later grow
into the observed Large Scale Structure. More
 interesting for us, are the short-distance, isocurvature
 ``miniclusters'', originally 
discussed by Hogan and Rees \cite{mini}, and
extensively studied by Kolb and Tkachev \cite{KT}. 
 These arise
because the axion field is random, from  one horizon to the
next, as the axion mass turns on before the QCD  PT.
This gives ${\cal O}(1)$ density fluctuations
on the QCD horizon scale, which are frozen in
the expanding radiation until matter-radiation equality,
then decouple from the Hubble expansion and collapse.

We want to estimate the mass of a minicluster. This
will depend on the  coherence length
and  energy density of  the axion field,
evaluated  as the axion mass turns on. 
The temperature dependence of the axion mass
is equivalent to  the topological susceptibility $\chi(T)$
of the QCD plasma, which has been widely studied. Here we consider
three possible behaviours for  $m^2(T)$
prior to the QCD phase transition.
In the interacting instanton  liquid of  \cite{WantzShellard},
the temperature-dependent axion mass can be written
\beq
m(T) = c \frac{\LQCD^2}{\fPQ} 
\left(\frac{\LQCD}{T}\right)^{n/2} \,,
\label{mT}
\eeq
with $\LQCD = 400$ MeV, 
and where $c = 4\times 10^{-4}$, $n = 6.68$.
(The zero-temperature  mass is obtained
for $n=0$, $c = 4 \times 10^{-2}$, and 
$m(T)$ should stop growing when it
reaches the zero-temperature mass.)
The instanton liquid gives a less-steep turn-on 
than  dilute instanton gas estimates~\cite{Turner,BHK},
and is the parametrisation used by KSS \cite{KSS}
to estimate that $m \sim 10^{-4}$ eV gives the
correct relic density of axion CDM  today.
A recent lattice-based estimate~\cite{Villadoro} suggests
that the axion mass could turn on more slowly, with
 $c \simeq 3 \times 10^{-3}$, $n \simeq 2$,
whereas the  lattice-based analysis of 
Kitano and Yamada \cite{KitanoYamada}
explored the possibility that the 
axion mass turns on exponentially
at the critical temperature $T_c \simeq 150$~MeV.


Once the axion mass reaches its zero temperature value $m$,  the
minicluster mass  can be estimated as
\begin{align}
  M_{mini} \sim V_{osc} \, m{(T=0)} \, n_{osc} \, E \,,
\end{align}
where $T_{osc}$ is the temperature when $m(T_{osc}) = 3H(T_{osc})$,
$V_{osc}$ and $n_{osc}$ are the volume of the horizon and the comoving axion
number density at that moment, and $E$ is an enhancement factor which
takes into account that the axion field is inhomogeneous so starts to
oscillate at different times in different horizons.
We take $V_{osc} \sim 1/[8H^3(T_{osc})]$ and
$n_{osc} \sim m(T_{osc}) \fPQ^2 \langle \theta_i^2 \rangle \approx
m(T_{osc}) \fPQ^2 \pi^2/3$.
It is reasonable to take the  minicluster volume to be
$V_{osc}$, because recall that the field was initially
random on much smaller scales, and is smoothed to the
horizon-scale by the dynamics of a massless scalar field. 
The enhancement factor $E$ arises because,
 in horizons where $a \sim \pi \fPQ$, the axion
potential does not have a $m^2 a^2$ form, and the
beginning of oscillations (with the associated
$1/R^3$ red-shifting of the energy density) is delayed.
Several studies~\cite{Turner,etal,BHK}
have  estimated $E \sim 2 \to 8$   by
numerically solving the equations of
motion\footnote{Notice that $E$ is the enhancement in the density
of overdense regions, whereas often the enhancement
in the average density is given.}.  
However it is unclear  in these analyses, whether the field
gradients are included in the equations of motion.
The gradients are explicitly included in
the analysis \cite{KT}, who find  that the
axion field could remain trapped at
$a \sim \pi \fPQ$ while the temperature dropped
by an order of magnitude. However, 
the $a \sim \pi \fPQ$ configurations
of this analysis may correspond to the
domain walls that form between strings,
after the axions get a mass, and whose decay
is studied by KSS~\cite{KSS}. We take
$E = 8$. (This may be large, but other
authors take the horizon volume to be $V= 1/H^3$.
We have $EV = 1/H^3$.) This leads to
\begin{align}
  \label{Mmini}
  M_{mini} \sim  \frac{\pi^2 m  \fPQ^2}{H^2(T_{osc})} \,.
\end{align}

With the parameters corresponding to the interacting
instanton liquid of  \cite{WantzShellard}, and
$\fPQ \simeq 6 \times 10^{10}$ GeV as found by
KSS, we find the axion starts to oscillate
around $T \sim 1 \to 2$ GeV, giving
\beq
M_{mini} \sim 3 \times 10^{-13} M_\odot \,.
\label{Mminin}
\eeq
This is similar to the value found in
\cite{Tkflens}, but smaller than the original
estimate of Hogan and Rees \cite{mini},
whose miniclusters formed at
$T = 100$ MeV in a CDM-dominated Universe.
In the case of  the  more gradual
turn-on of the axion mass advocated in \cite{Villadoro},
the axion starts to oscillate  sooner
($T_{osc} \sim$ few GeV), so the miniclusters
could be an order of magnitude smaller. 
If the axion mass turns on exponentially at
the QCD phase transition, as discussed in
\cite{KitanoYamada},  then it is possible
that the axion would  contribute to much CDM. 
Requiring that  the observed dark matter density is
due to axions  in this scenario, the
axion could
start to oscillate at $T \sim \LQCD$,
 giving miniclusters an order of magnitude 
or so larger. 

Interestingly, $M_{mini}$ of eqn~(\ref{Mminin})  is  not so different from
the maximum mass of a stable drop 
estimated in eqn~(\ref{Mdrop}).
This appears to be an accident: the maximum mass of
a drop is determined by balancing the
kinetic pressure against the gravitational
and self-interaction forces --- whereas the
mass of a minicluster is determined by the volume
of the horizon when the axion starts to
oscillate.  To see why these  are similar, recall that the
axion starts to oscillate when $3H(T_{osc}) \sim m(T_{osc})$. Replacing
$25 m \fPQ \sim  \LQCD^2$  in eqn~(\ref{mT}),
gives 
$$
M_{mini} \sim \frac{\pi^3 m  \fPQ^2}{ H (T_{osc})}  \frac{1}{ m (T_{osc})}
\sim  \pi^3 m  \fPQ^2 \frac{\mpl}{ 15  T^2}  \frac{ 4T^{2}}
{ m ^2 \fPQ }
\left(\frac{ T}
{25 m \fPQ}\right)^{ \frac{n-4}{4}} 
\sim  \frac{\fPQ \mpl}{m} \frac{8 T^{1.34}}{ \LQCD^{1.34} }
$$ 
where the last estimate used  $n/2 = 3.34$, for
which  the last
fraction is $\sim 50$.
The estimates of the minicluster and drop masses,
given in eqns (\ref{Mminin})  and (\ref{Mdrop}), are
both  very  uncertain. 
In both cases, formation is a dynamical process 
which  should generate a spectrum  of masses. So
we take these estimates to have an uncertainty
of at least an order of magnitude, and conclude
that  the average minicluster mass could be
of order the maximum drop mass, or a hundred
times larger.

\subsection{ Speculations on how to make Andromeda}
\label{Andromeda}

We can now speculate on how axion dark matter could make
the Andromeda galaxy, in the case where the Peccei-Quinn phase transition
occurs after inflation. 

After the QCD phase transition, there are two contributions
to axion dark matter: the misalignment field, and the incoherent
distribution of  modes produced by  strings. It is convenient
to refer to this phase space distribution of modes as ``particles'',
to distinguish  it  from the misalignment field (distributions
 of classical waves  evolve in a way very similar to
distributions of particles~\cite{AB02}). Once the axion
mass has settled to its zero-temperature value,
the particles are non-relativistic, with
velocity $\sim H(\LQCD)/m \sim  10^{-5} c$
(for $m= 10^{-4}$ eV). Despite their small mass, these
particles should be sufficiently ``cold''
to virialise and form a  galactic halo of
particles with $v\sim 0.001c$.

There are two types of
density perturbations that can arise
in     the field  and
in the particles.    On large scales,
they both inherited from the surrounding radiation,
  the scale-free, adiabatic density fluctuations
produced during inflation.
Then, due to the dynamics at the QCD phase transition, 
the field has ${\cal O}(1)$ density fluctuations on
the QCD horizon scale ($\sim 0.03$~pc today),  
which give a white noise spectrum ($\delta M/M \sim \sqrt{M_{mini}/M}$)
 of density fluctuations on small scales.
 These  are the miniclusters. The particles might also
have a similar small-scale spectrum. These
 density fluctuations
are frozen into the radiation plasma until they
dominate the radiation density (see {\it e.g.}~\cite{Efstathiou}). 
At    matter-radiation
equality,
the large-scale fluctuation which will eventually become
Andromeda, which has an initial amplitude $ \delta \rho/ \rho \lsim 10^{-4}$,
can start to  grow. Meanwhile,  the  short-distance  
${\cal O}(1)$ inhomogeneities, such as miniclusters,
 decouple from the Hubble flow and collapse. 

We now focus on the smale-scale inhomogeneities in
the field.
A realistic and  accurate calculation of  minicluster formation
would give a spectrum: the
number density of miniclusters as a function of
their mass. Those with mass less than the
maximum  allowed for drops should initially  collapse to axion drops.
Hogan and Rees \cite{mini} speculate that 
miniclusters undergo hierarchical clustering,
however it is unclear to us how this would occur:
do small drops amalgamate to form larger drops,
or  do the drops cluster like dark matter particles?

In the case that a minicluster exceeds the maximum drop
mass,  the virial theorem suggests that the
miniclusters fragment into  stable drops.
The point is that the virial condition implies that
the (negative) gravitational energy  gained in collapse
should be  compensated by kinetic energy, so
one can anticipate that the minicluster fragments
into smaller field configurations, in the presence
of steep field gradients. This would of course
require numerical verification.  The alternative
is that the minicluster  collapses to a black hole.
However,  it is often argued that  ${\cal O}(1)$
density fluctuations should be of horizon-scale in order  
to collapse to black holes \cite{Karsten}, which is not
the case for miniclusters.

In this paper, we do not consider small-scale
inhomogeneities in the  density of axions produced by strings.
The string network is very inhomogeneous,  but decays to
relativistic axion particles~\cite{KSS}, which  become
non-relativistic as the  temperature-dependent axion mass 
increases towards its zero-temperature value. 
The degree to which  the particle  density becomes
smooth  by free-streaming,  prior to  the axions
becoming non-relativistic, is unclear (to us).
Kolb and Tkachev explored this question \cite{Tkflens},
using a dilute instanton gas approximation for
$m(T)$, and argue that the axions from strings
also  have ${\cal O} (1)$ density fluctuations
on the QCD horizon scale.  This corresponds to
the  original analysis of 
Hogan and Rees  \cite{mini}, where the miniclusters
were assumed to be composed of axion particles.
Hogan and Rees  estimate that the cores of miniclusters
could survive  the hierarchical merging of miniclusters
and the galaxy formation process, but  that
a significant fraction of axion particles would be
thrown off,  and would today  contribute a  smooth
halo density of axion particles, as could for instance
be detected by axion dark matter search experiments
such as ADMX \cite{ADMX}. 

On the other hand, if a significant fraction of  
axion dark matter is in the form of
drops, this would reduce the signal in direct detection.
Denoting
the mass fraction of axion drops to the
smooth halo component by $f_{drop}$, we expect a number density of
drops with mass $10^{-13}\,M_\odot$ of order $f_{drop} \times
10^{-44}$~cm$^{-3}$, which implies about $f_{drop} \times 10^{-5}$
drops in a volume of (1~AU)$^3$ and a drop flux on Earth of $f_{drop} \times
10^{-34}\,\rm cm^{-2} \, s^{-1}$. (We have assumed a local dark matter
density of 0.4~GeV~cm$^{-3}$ and a drop velocity in the galactic halo
of $10^{-3}\,c$.) Hence, ADMX would see only the smooth component of
the halo, which will be reduced by a factor $(1-f_{drop})$.


\section{Observational bounds}
\label{secC}

The observational signatures of ``macroscopic'' dark matter objects,
with masses  from grams to 
several solar masses  were recently
compiled in \cite{macro}. The case of primordial black
holes  is reviewed in \cite{PBH};  while black holes
created with $M \lsim 10^{-18} M_\odot$ would evaporate
in the lifetime of the Universe, there is a window \cite{Griest2}
$ 10^{-13} M_\odot \lsim M_{PBH} \lsim  10^{-9} M_\odot$
where they could constitute the dark matter 
(the lower bound is from femtolensing, the upper bound
from microlensing).   This window is interesting, because
in the case that large miniclusters 
collapsed to black holes,  they would be in this allowed range. 
Kolb and Tkachev \cite{Tkflens} discussed the
sensitivity of femto- and pico-lensing experiments
axion miniclusters. Zurek etal \cite{Zurek:2006sy} consider astrophysical effects
of miniclusters in the wide range of $10^{-12} \to 10^4 \, M_\odot$.
Here we suppose that axion field dark matter is in the form of drops,
with $M \lsim 10^{-13} \to 10^{-12} M_\odot$ and  review the  femtolensing bound.
Other possible  constraints are listed. 

\subsection{Femtolensing}

Micro-lensing is the familiar idea of  watching
nearby stars  ({\it e.g.} in the LMC), in the hope
of observing  an increase in their light  due to a compact 
halo object crossing the line of sight.  Femtolensing
\cite{GouldGRB}
uses Gamma Ray Bursters (GRBs) as sources, which are at cosmological distances,
and most of which only last for a few seconds. 
The  lensing objects are therefore distributed
in intervening galaxies and intergalactic space.
And rather than looking for an amplification in
light signal, one looks for the interference
between light that took two different paths
round the lensing object: the time delay
between the two paths is the same for
photons of different energies, so one
looks for oscillations in the energy
spectrum of GRBs. 

Femtolensing   is  an idea of
 Gould \cite{GouldGRB}, that  lensing by 
dark objects with  $10^{-16} M_\odot \lsim M  \lsim  10^{-13} M_\odot$ 
could give interference patterns in the
energy spectrum of GRBs.  
A  bound  based on BATSE data
\cite{Marani}, could  exclude $\Omega \sim 0.2$
for the mass range $10^{-16} \to 10^{-13} M_\odot$, 
and these authors  also estimated picolensing
bounds,  and found 
a 1 $\sigma$ sensitivity   to $\Omega \sim 1$ 
of  compact objects
in the mass range  $10^{-12.5} M_\odot \to 10^{-9} M_\odot$.
In a  careful
and dedicated analysis of FERMI data, Barnacka etal  \cite{BGM}
focussed on GRBs at measured redshift,
and were able to exclude  $\Omega > 0.03 $ 
in  compact objects of   mass  
between   $5\times 10^{-17} \to 5\times 10^{-15} M_\odot $,
which is somewhat below the maximum drop mass
estimated in eqn~(\ref{Mdrop}).

The exclusion  of   \cite{BGM} assumes
that the GRB can be treated as point source.
Otherwise, various  photons emitted by
the GRB could have different time delays
in their paths around the lens
(because they come from different locations on
the source), and  the oscillations
in the intensity-summed-over-photons
could be washed out~\cite{GouldGRB}.
The GRB  can be treated as point source
provided that its ``size'', projected onto
the lens plane, is smaller than the
Einstein  radius $r_E = 2\sqrt{G_N M(D_{OL} D_{LS}/D_{OS})}$,
where the distances $D$ are between Observer, Source
and Lens. It is unclear, from the
estimates  in \cite{BGM}, that this condition
is satisfied (such doubts
were expressed in \cite{PaniLoeb}). Furthermore, the GRB size
estimates in \cite{BL} are one or two
orders of magnitude larger than
those in \cite{BGM}. It would be helpful
if the status of these constraints was confirmed
by expert authors.

\subsection{Other constraints?}
\label{ssec:Other}

Axion drops  as dark matter could have many other observable 
consequences, due to their interactions with the CMB,
 magnetic fields, or other astrophysical objects, due
to their passage in our local area,   and so on.
Some possibilities are listed here. 
Questions which arise,  for some of these constraints,
are   whether the drop accretes baryons (see {\it e.g.}~\cite{Zurek:2006sy}), 
and how it  interacts with photons (discussed
in \cite{Espriu}).

\ben
\item Carr and Sakkelariadou \cite{CS}
considered dynamical constraints on compact objects, 
which could disrupt the structures we see.
They  expect that  compact objects   in the mass range
$10^{-18} \to 10^{-11} M_\odot$  could ressemble comets.
 From the non-observation of interstellar comets 
in the past 300 years, they  impose  that no compact
objects (CO) passed through a disk of radius  one 
earth-sun distance (astronomical unit= AU)  in
300 years, which implies,  following \cite{Hills}:
\bea
M > \frac{\Omega_{\rm CO}}{\Omega_{\rm DM}} 10^{-13} M_\odot \,.
\label{HillCS}
\eea
So for drop masses  $\gsim$ few $\times 10^{-14} M_\odot $,
this observation does not pose a constraint.
Whether there is a bound on smaller drops would require study,
to determine whether they shine
like baryonic comets.

\item  There are constraints on DM-photon interactions from
 CMB observations, for instance as given in \cite{JulienCeline} ---
it would be interesting to understand if these
apply to axion drops? In \cite{macro}, it is argued
that  macroscopic compact objects  have a geometric
cross-section with photons, and  can be subject to
the same ``collisional damping'' (Silk damping) constraints
as particle dark matter.    Whether this is
the case for drops  might depend on 
whether they accumulate baryons.

\item Do the drops evaporate due to self-interactions?

The  rate at which
four axions from the condensate (field)  could
evaporate into two particles with energy
$\sim 2m$ is estimated\footnote{This
can be simply obtained  as the
$2\to 2$ scattering rate with an
effective four-point coupling
$m^2 a^2/\fPQ^6$.}
 in \cite{PWW} as
$ \sim m^3 a^6/ \fPQ^4$. For $a/\fPQ \sim \fPQ/\mpl$,
as obtained in the axion drop, this evaporation
timescale is much longer than the age of the Universe.

The decay of axion drops due to emission of real axions because of the
the self-interaction term via a $3a \to a$ process has been studied in
\cite{Eby:2015hyx}. This process becomes kinematically allowed when the
whole axion drop balances momentum. It has been found that if
configurations close to the maximal possible total mass of the drop
are considered, the decay becomes relevant for axion masses around
$\lesssim 10^{-8}$~eV, but should be unimportant for $m\sim
10^{-4}$~eV, as implied by $\fPQ \sim 10^{11}$~GeV.

Axion particles in the galactic halo 
(originally   produced by
string decay), could  scatter axions out of
the drops. The rate for this process was estimated
in \cite{axions2} to be $\sim n_a  m^2/\fPQ^4 \times f_{\rm BE}$,
where $f_{\rm BE}$ is a Bose enhancement factor
that accounts for the high occupation number of
the axion particles, which can
locally be  estimated as  
$f_{\rm BE} \sim 0.3$~GeV/($m^4v^3$cm$^3$) $\sim 10^{20}$
(for $v\sim 0.001c$ the local virial velocity).
The fraction of axions scattered
out of drops in the age of the Universe 
 $\tau_U$ is therefore
$\sim (m f_{\rm BE}/\mpl) \ll 1$.


\item One can ask what  happens if a drop meets an
ordinary star, a white dwarf,  a neutron star,  or a
black hole.  Collisions of 
axion drops  with white dwarfs  and neutron stars
\cite{Iwazaki}
have been proposed as a source for GRBs, as
well as to explain other anomalies. 
However the energy released in
the collision of a drop with a neutron 
star is controversial \cite{Barranco:2012ur}
(as is also the case for the
interactions of primordial black holes
with neutron stars \cite{PaniLoeb,Tinyakov,CPT}).

\item  The ``explosion''  of axion drops was recently proposed
as a possible source for Fast Radio Bursts \cite{TkachevFRB}.

\een


\section{ Summary}
\label{sum}

Dark matter composed of the QCD axion can be produced
either by the misalignment mechanism, giving rise
to an axion field, or by the decay of strings,
which produces a distribution of axion modes/particles.
We consider stable axion field configurations,
held together by gravity and self-interactions, 
and confirm that they have 
typical dimensions of 100~km and a mass scale $M \sim 10^{-13}
M_\odot \sim 10^{-7} M_\oplus \sim 10^{17}$~kg.
The maximum mass of the drop arises  from
balancing the inward gravity and self-interaction pressures against
kinetic gradient pressure.

We allowed the  axion drops to rotate, and 
found that the maximum mass can increase by about an order of
magnitude. This result is estimated analytically  in section
\ref{secA}  using the  virial
theorem, and obtained  in section~\ref{sec:num} from
numerical solutions of the classical field
equations. In both cases we have assumed that the axion field in the
drop is proportional to a single spherical harmonic $Y^l_m(\vartheta,
\varphi)$. In realistic situations one may expect
that the field is a general linear combination of several $Y^l_m$.
Nevertheless, our simplified ansatz indicates that modestly
rotating drops ($l \lesssim $~few) may be somewhat heavier than
non-rotating configurations.

In this work, we looked for stable solutions, but did not
study  the dynamical process of drop formation
(which  would depend on the earlier cosmological evolution).
Nonetheless,  in section \ref{secB}, we speculate
on how axion drops could arise if
the Peccei-Quinn phase transition occurs
after inflation. In this scenario, 
the misalignment axions are only a component of the dark
matter, and have ${\cal O}(1)$ inhomogeneities on
the horizon scale of the QCD phase transition which
are refered to as miniclusters.
The estimated mass of miniclusters 
is slightly larger than  the maximum
mass of stable axion drops, so we envisage that
the miniclusters could fragment into drops.
On the other hand, if the Peccei-Quinn phase transition occurs
before inflation, then the misalignment axions
compose all the axion dark matter, but  there
are no miniclusters, and it is an open
question whether  large-scale density fluctuations
would fragment on small scales into drops.

In section \ref{secC}, we reviewed  observational constraints on  dark
matter in the form of asteroid sized objects, and it appears that
dark matter halos  made of axion drops could be
consistent with observations. Interesting constraints close to the
relevant mass range are obtained by femtolensing. Several other
potential constraints depend  whether axion drops accrete
baryons, which we did not study.  

It is interesting to speculate on the
implications  of axion drops for direct detection experiments, 
such as  ADMX~\cite{ADMX}.
Recall that in the cosmological
scenario where the Peccei-Quinn phase transition occurs
after inflation, current numerical simulations  \cite{KSS}
suggest that $\sim$ 75\% of axion dark matter
is  composed of particles produced  by strings. 
As reviewed at the end of section \ref{Andromeda},
these axion particles could provide  a smooth
halo  component, as  searched for by ADMX.  However,
the dark matter fraction stored in drops would be
hidden from axion direct detection experiments.

\subsection*{Acknowledgements}
SD thanks Georg Raffelt  for useful, pleasant, and essential discussions.


\begin{appendix}

\section{Non-relativistic approximation}
\label{app:NR}

In general a scalar field coupled to gravity is described by the following action
\begin{equation}
  S = \int d^4x \sqrt{-g} \left[ \frac{\mathcal{R}}{16\pi G_N} + \mathcal{L} \right] \,,
\end{equation}
where $g \equiv \text{det}(g_{\mu\nu})$ is the determinant of the
metric and $\mathcal{R}$ is the Ricci scalar. We consider here the Langrange
density of the relativistic real axion field $a$ of the form
\begin{align}
  \mathcal{L} = - \frac12 g^{\mu\nu}(\partial_\mu a)(\partial_\nu a) - V(a) \,,\qquad
  V(a) = \frac12 m^2 a^2 - \frac{1}{4!} \lambda a^4 \,, 
\end{align}
with the dimensionless quartic coupling $\lambda = m^2/\fPQ^2$. When the action is
extremized with respect to the metric $g_{\mu\nu}$ and the field $a$
one obtaines the coupled Klein-Gordon and Einstein equations.
Here we are going to derive the non-relativistic limit, where all
velocities are small compared to $c$ and energies small compared to
$m$. We follow closely Ref.~\cite{NambuSasaki}. 

For the metric we take the Newtonian ansatz with
\begin{equation}
  g_{00} = -(1+2 V_N)\,,\quad g_{i0} = 0 \,,\quad g_{ij} = (1-2 V_N)\delta_{ij} \,,
\end{equation}
with the Newtonian potential $ V_N(\vec{x}) \ll 1$ and we neglect the
explicit time dependence of $ V_N$. Hence, we have $\sqrt{-g} \approx
1-2 V_N$.  To leading order in $ V_N$ the Ricci scalar is given by $\mathcal{R}
= - 2 (\partial_i V_N)^2$ (see for instance \cite{Dodelson} eqn~5.17
with metric definition 4.9), and we obtain for the action
\begin{equation}
  S = \int d^4x \left[ -\frac{(\partial_i V_N)^2}{8\pi G_N} +
    \frac{1-4 V_N}{2} {\dot a}^2 -  \frac12 (\partial_i a)^2 - (1-2 V_N) V(a) \right] \,,
\end{equation}
with dot denoting time derivative. In order to take the
non-relativistic limit for the axion we write the real relativistic
field $a$ in terms of a complex non-relativistic field $\phi$ as
in eqn~\eqref{NRa}.
%
The non-relativistic approximation assumes that $\phi$ varies only slowly on time scales
$1/m$:
\begin{equation}\label{eq:nr-approx}
  \dot \phi \ll m \phi \,,\qquad \ddot \phi \ll m \dot \phi \,,
\end{equation}
and all factors containg exponentials $e^{\pm i mt}$ average to zero
when integrated over $t$ in the action. Note that with the ansatz for
$\phi$ from eqn~\eqref{eq:psi-ansatz} the above approximation just
means $E \ll m$, which has been confirmed to hold for our solutions,
see eqn~\eqref{eq:E}. Under these assumptions we find
\begin{align}
  {\dot a}^2 &\to i(\dot\phi \phi^* - {\dot \phi}^*\phi ) + m\phi\phi^* \,,\qquad
  (\partial_ia)^2 \to \frac{1}{m} (\partial_i \phi) (\partial_i \phi^*) \,,\qquad
  a^2 \to \frac{1}{m}\phi\phi^* \,,\qquad
  a^4 \to \frac{3}{2m^2}(\phi\phi^*)^2 \,.
\end{align}
In the expression for $ {\dot a}^2$ we have neglected a term ${\dot
 \phi}^* \dot\phi / m$ but kept the ones with only one derivative.
Then we obtain the action in terms of the fields $ V_N, \phi, \phi^*$:
\begin{equation} \label{eq:action}
  S = \int d^4x \left[ -\frac{(\partial_i V_N)^2}{8\pi G_N} +
    \frac{i}{2}(\dot\phi \phi^* - {\dot \phi}^*\phi ) - m V_N \phi\phi^* -
    \frac{1}{2m}(\partial_i \phi) (\partial_i \phi^*) +
    \frac{(1-2 V_N)\lambda}{16 m^2}(\phi\phi^*)^2 \right]\,.
\end{equation}
We have neglected terms of order $ V_N \dot\phi \phi^*$ compared to
$m V_N \phi\phi^*$ according to eqn~\eqref{eq:nr-approx}.  Note that
the large terms of order $m\phi\phi^*$ cancel, as a consequence of
factoring out the fast oscillations with frequency $m$, which
basically reduces the Klein-Gordon equation to the Schr\"odinger
equation.  Applying Euler-Lagrange equations for $\phi^*$ to the
action \eqref{eq:action} we find the GP equation,
eqn~\eqref{GP}, with $g = \lambda/(8m^2) = 1/(8\fPQ^2)$,
see eqn~\eqref{Veta}, and we neglect the gravitational potential
$V_N \ll 1$ in the term proportional to $\lambda$.

The Euler-Lagrange equations for $ V_N$ lead to the following equation:
\begin{align}
  \Delta V_N = 4\pi G_N \left(m + \frac{\lambda}{8m^2} \phi\phi^* \right) \phi\phi^* \,.
\end{align}
Apart from the second term in the bracket this is the Poisson equation
eqn~\eqref{poisson}.  Let
us use the results of section~\ref{sec:num} to estimate the relative
size of the two terms in the bracket:
\begin{align}
  \frac{\lambda}{8m^3} \phi\phi^* = \frac{1}{8} \frac{R^2 }{m f_{\rm PQ}^2}  =
  64\pi^2 \frac{f_{\rm PQ}^2}{M_{\rm Pl}^2} {\tilde R}^2 \sim 10^{-11} \,,
\end{align}
where we have used $\lambda = m^2/f_{\rm PQ}^2$ and
eqn~\eqref{eq:tilde-R} to estimate $R^2$.  Hence, the term
proportional to $\lambda$ can be savely neglected and we recover the
Poisson equation.

Let us consider the Schwarzschild radius $r_s =  2 G_N M$. 
Using eqs~\eqref{eq:tilde-R} and \eqref{eq:tilde-M} we obtain
\begin{equation}
  \frac{r}{r_s} = \frac{1}{32\pi}\frac{\mpl^2}{f_{\rm PQ}^2}\frac{\tilde r}{\tilde M}
  \approx 10^{14} \frac{\tilde r}{\tilde M} 
\left( \frac{f_{\rm PQ}}{10^{11} \, \rm GeV} \right)^{-2} \,.  
\end{equation}
Hence, general relativistic effects are small and the Newtonian
treatment of gravity is justified.


\end{appendix}


\end{document}